\documentclass[apjl]{emulateapj}
\usepackage{amsmath}
\usepackage{color}
\usepackage{graphicx}
\newcommand{\simpl}{{\sc simpl~}}
\newcommand{\simplone}{{\sc simpl-1~}}
\newcommand{\simpltwo}{{\sc simpl-2~}}
\newcommand{\powerlaw}{{\sc powerlaw~}}
\newcommand{\comptt}{{\sc compTT~}}
\newcommand{\kerrbb}{{\sc kerrbb~}}
\newcommand{\kerrbbtwo}{{\sc kerrbb2~}}
\newcommand{\bhspec}{{\sc bhspec~}}

\newcommand{\simplb}{{\sc simpl}}
\newcommand{\simploneb}{{\sc simpl-1}}
\newcommand{\simpltwob}{{\sc simpl-2}}
\newcommand{\powerlawb}{{\sc powerlaw}}
\newcommand{\compttb}{{\sc compTT}}

\newcommand{\kerrbbtwob}{{\sc kerrbb2}}

\newcommand{\msun}{\rm M_{\sun}}

\newcommand{\rchinu}{\chi^{2}/\nu}
\newcommand{\fsc}{f_{\rm SC}}
\newcommand{\nh}{N_{\rm H}}

\newcommand{\kpc}{\rm kpc}

\newcommand{\cm}{\rm cm}

\newcommand{\rin}{R_{\rm in}}

\newcommand{\ledd}{L_{\rm Edd}}
\newcommand{\lledd}{L_{D}/L_{\rm Edd}}

\newcommand{\rxte}{{\it RXTE~}}

\newcommand{\rxteb}{{\it RXTE}}

\shorttitle{{\sc BH Spin Beyond the TD State}}
\shortauthors{Steiner et al.}

\begin{document} 

\title{Measuring Black Hole Spin via the X-ray Continuum Fitting Method: Beyond the Thermal Dominant State} 

\author{James F.\ Steiner\altaffilmark{1}, Jeffrey E.\
        McClintock\altaffilmark{1}, 
        Ronald A.\ Remillard\altaffilmark{2},
        Ramesh Narayan\altaffilmark{1},
        and Lijun Gou\altaffilmark{1}}

\altaffiltext{1}{Harvard-Smithsonian Center for Astrophysics,60 Garden Street,
Cambridge, MA 02138.} \altaffiltext{2}{MIT Kavli Institute for Astrophysics and Space Research, MIT, 70 Vassar Street, Cambridge, MA 02139.}

\email{jsteiner@cfa.harvard.edu}

\begin{abstract} 
  All prior work on measuring the spins of stellar-mass black holes
  via the X-ray continuum-fitting method has relied on the use of
  weakly-Comptonized spectra obtained in the thermal dominant state.
  Using a self-consistent Comptonization model, we show that one can
  analyze spectra that exhibit strong power-law components and obtain
  values of the inner disk radius, and hence spin, that are consistent
  with those obtained in the thermal dominant state.  Specifically, we
  analyze many {\it RXTE} spectra of two black hole transients,
  H1743--322 and XTE J1550--564, and we demonstrate that the radius of
  the inner edge of the accretion disk remains constant to within a
  few percent as the strength of the Comptonized component increases
  by an order of magnitude, i.e., as the fraction of the thermal seed
  photons that are scattered approaches 25\%.  We conclude that the
  continuum-fitting method can be applied to a much wider body of data
  than previously thought possible, and to sources that have never
  been observed to enter the thermal dominant state (e.g., Cyg X--1).
\end{abstract}


\section{Introduction}\label{section:Intro}

Black holes (BHs) are completely described by only three quantities:
mass, charge, and spin.  In astrophysical settings, any net charge
will rapidly neutralize, with the result that a stellar-mass BH is
specified by just its mass and spin.  BH spin is commonly expressed in
terms of the dimensionless parameter $a_* \equiv cJ/GM^2$ with $|a_*|
\le 1$, where $M$ and $J$ are respectively the BH mass and angular
momentum, and $c$ and $G$ are the speed of light and Newton's
constant.
While mass measurements of stellar-mass BHs have been made for
decades, the first spin measurements have been achieved only during
the past three years (\citealt{Shafee_spin, spin_1915, spin_m33,
  Gou_2009, Miller_2009}, and references therein).  Meanwhile, the
spins of supermassive BHs have also been measured
\citep{Brenneman_Reynolds, Miniutti_2007}.  The only two methods
presently available to measure BH spin are via modeling the thermal
continuum spectrum of a BH accretion disk, as pioneered by
\citet{Zhang}, or by modeling the profile of a relativistically
broadened Fe K fluorescence line, as demonstrated by \citet{Tanaka_1995}.

Spin is measured by estimating the inner radius of the accretion disk
$R_{\rm in}$.  One identifies $R_{\rm in}$ with the radius of the
innermost stable circular orbit $R_{\rm ISCO}$, which is dictated by
general relativity.  $R_{\rm ISCO}/M$ is a monotonic function of
$a_*$, decreasing from $6G/c^2$ to $1G/c^2$ as spin increases from
$a_*=0$ to $a_*=1$ \citep{Shapiro_Teukolsky}.  This relationship
between $a_*$ and $R_{\rm ISCO}$ is the foundation of both
methods of measuring spin.

In the continuum-fitting (CF) method, one determines $R_{\rm ISCO}$ by
modeling the X-ray continuum spectrum, focusing on the thermal
accretion-disk component.
The observables are flux, temperature, distance $D$, inclination $i$,
and mass $M$.  To obtain reliable values of spin, it is essential to
have accurate estimates for $M$, $i$ and $D$, which are typically
derived from optical data.

The CF method has been applied only to spectral data obtained in the
thermal dominant (TD) state (or very recently to a near-TD
intermediate state; \citealt{Gou_2009}).  The TD state is chiefly
characterized by the dominance of the soft, thermal disk component of
emission.  (For a measure of this dominance and a review of BH states,
see \citealt{RM06}.)  The CF method has never been applied to the more
Comptonized steep power law (SPL) state, which is characterized by the
coexistence of a strong power-law component with photon index $\Gamma
> 2.4$ and a significant thermal component.  Most models for the SPL
state invoke Compton up-scattering of thermal seed photons by coronal
electrons as the mechanism that generates the power law.  Herein, we
employ a self-consistent Comptonized accretion-disk model that yields
values of $R_{\rm in}$ for SPL-state spectra that are consistent with
those obtained for TD-state spectra.  This result greatly increases
the reach of the CF method, allowing one to obtain reliable
measurements of spin for a much wider body of data than previously
supposed, and for sources that do not enter the TD state (e.g., Cyg
X-1).  Moreover, the success of this model supports the widely-held
assumption that Comptonization is the mechanism which generates the
observed high-energy power law component in SPL- and TD-state spectra.

Our full model of a Comptonized accretion disk is a convolution of the
relativistic thin accretion-disk model {\sc kerrbb2}
\citep{KERRBB,spin_1915} and {\sc simpl}, an empirical model that
convolves a Comptonization Green's function with an arbitrary seed
photon spectrum \citep{Steiner_simpl}.  Both models are implemented in
XSPEC \citep{XSPEC}.  \simplb, with only two parameters, ensures
photon conservation and self-consistently generates the power-law
component of the spectrum of an accreting BH using the accretion-disk
component as input.

We have chosen to apply our spectral model to the two bright transient
X-ray sources H1743--322 (hereafter H1743), which we feature, and
XTE~J1550--564 (hereafter J1550).  Both transients are sources of
large-scale relativistic jets and high-frequency QPOs (\citealt{RM06},
and references therein).  For a detailed comparison of the spectral
and timing characteristics of these very similar transients during
their principal outbursts,
see \citet{JEM_H1743_2007}.  Presently, the distance to J1550 is
poorly constrained (see \citealt{Orosz_2002}), and no useful distance
estimate or dynamical information whatsoever is available for the
black hole candidate H1743.  Consequently, we cannot yet accurately
estimate the spins of these black holes.  In this work, we adopt
fiducial values of $M$, $i$ and $D$.  Of course, $R_{\rm in}$ (and
$a_*$) depend strongly on these fiducial values.  However, as we show
in \S\ref{subsec:dynmod}, for any reasonable range of these input
parameters, the dependence of $R_{\rm in}$ on luminosity or on time
during the outburst cycle is slight, which is an important conclusion
of this work.

We show that the very widely-used additive XSPEC models of
Comptonization, namely the empirical model \powerlaw and the physical
model \comptt (\citealt{COMPTT}; \S\ref{section:Res}), are inadequate
for extracting measurements of spin from spectra with substantial
power-law components.  A self-consistent model such as {\sc simpl} is
required.

\section{Observations \& Analysis} \label{section:Obs} 

We apply the model described below to the full archive of spectral
data for the 2003 outburst of H1743 (the most intense observed for
this source) and for all five outburst cycles of J1550 obtained using
the Rossi X-ray Timing Explorer's (\rxteb's) Proportional Counter
Array (PCA; \citealt{RXTE}).  We rely solely upon ``standard 2''
spectra obtained using the PCU-2 module, \rxteb's best-calibrated
detector.  All spectra have been binned into approximately half-day
intervals, background subtracted, and have typical exposure times
$\sim3000\;$s.  For the first 5 weeks of PCA observations (through
2003 May 1 UT) the detector was pointed $0.32^{\circ}$ from H1743.  We
have corrected the fluxes to full collimator transmission assuming a
triangular response with FWHM = $1^{\circ}$.  We applied similar
collimator corrections ($\approx 0.1^{\circ}-0.3^{\circ}$) to three
observations of J1550 performed on 1998 September 7--8 and 1999
January 5 UT.

A 1\% systematic error has been included over all channels to account
for uncertainties in the response of the detector (details on \rxteb's
calibration can be found in \citealt{RXTE_cal}).  As in our earlier
work (e.g., \citealt{spin_1915}), we have corrected for detector dead
time while using contemporaneous Crab observations and the canonical
Crab spectrum of \citet{Toor_seward} in order to calibrate the PCA
effective area.  The resultant pulse-height spectra are analyzed from
$2.8 - 25$ keV using XSPEC v12.5.0.

In XSPEC, the model we employ is {\sc
  phabs}(\simplb$\otimes$\kerrbbtwob), where {\sc phabs} is a
widely-used model of low-energy photoabsorption.  \simpl redirects
photons from the seed distribution, described here by the
accretion-disk model \kerrbbtwob, into a Compton power law.  Like
\powerlawb, \simpl has just two parameters: (1) the fraction of seed
photons $\fsc$ scattered into the power law, and (2) the photon
power-law index $\Gamma$.  \simpl does not incorporate higher-order
effects such as geometry-dependent scattering or reflection.  The
relativistic disk model \kerrbbtwo similarly has two fit parameters:
(1) the spin parameter $a_*$, which we express equivalently in terms
of $R_{\rm in}$ (\S\ref{section:Intro}), and (2) the mass accretion
rate $\dot M$.  From these two parameters we compute the
Eddington-scaled disk luminosity, $L_{D}(a_*,\dot M)/L_{\rm Edd}$,
where $L_{D}$ is the luminosity of the seed photons and $L_{\rm Edd}
\approx 1.3 \times 10^{38} M/\msun$~erg s$^{-1}$
(\citealt{Shapiro_Teukolsky}).  The low-energy cutoff is parameterized
in the {\sc phabs} component by the column density $\nh$, which we fix
at 2.2$\times 10^{22}~\cm^{-2}$ for H1743 and 8$\times
10^{21}~\cm^{-2}$ for J1550~\citep{JEM_H1743_2007,Miller_2003_j1550}.
We include an additional model component to account for
disk-reflection using the XSPEC model {\sc smedge} for J1550, which
was not required for H1743.

In our analyses described in
\S\S\ref{subsec:selection},\ref{subsec:compare}, we adopt the
following dynamical model parameters: For H1743, $M=10~\msun$,
$i=60\degr$ and $D=9.5~\kpc$; and for J1550, $M=10~\msun$,
$i=70\degr$, and $D=5~\kpc$.  The values for H1743 are chosen
arbitrarily to place the maximum outburst disk luminosity at
$L_{D}/\ledd \approx 0.7$, and those for J1550 are round numbers based
on the model described in \citet{Orosz_2002}.  In
\S\ref{subsec:dynmod}, we allow $i$ and $D$ to vary and consider six
disparate dynamical models.

For H1743 and J1550 we only select data over an order of magnitude in
luminosity, between $0.05 < L_{D}/\ledd < 0.5$ for the values of $M$,
$i$, and $D$ given above.  This intermediate range in luminosity is
chosen in order to eliminate both hard-state spectra that have little
or no detectable thermal component and high-luminosity data for which
the thin-disk approximation likely no longer applies.  Further
requiring goodness-of-fit ($\rchinu$) $<$ 2 and that the inner radius
is well-determined ($\rin/\Delta\rin>5$, where $\Delta\rin$ is the
$1\sigma$ statistical uncertainty on $\rin$) leaves us with a total of
117 spectra for H1743 and 151 spectra for J1550.

We include for \kerrbbtwo the effects of limb darkening and returning
radiation and set the torque at $R_{\rm in}$ to zero (e.g,
\citealt{spin_1915}), and for the dimensionless viscosity parameter we
adopt $\alpha=0.01$.
(Our results in the following section are modestly affected if one
instead uses $\alpha=0.1$: $\rin$ is increased by $\approx 5\%$ and
becomes weakly dependent on luminosity, increasing by $\lesssim 10\%$
for an order of magnitude increase in $L_{D}$.)  A color correction
resulting from spectral hardening in the disk atmosphere is internally
calculated for \kerrbbtwo using models \kerrbb and \bhspec
\citep{BHSPEC} as described in \citet{spin_1915}.  The
upscattering-only implementation of {\sc simpl}, known as {\sc
  simpl-1}, was used exclusively throughout unless otherwise noted.
Larger values of $\fsc$ are obtained using the double-sided scattering
kernel \simpltwo (see Table 1), but $\rin$ and the other fit
parameters are completely unaffected by the choice of kernel.

\section{Results}\label{section:Res}

\subsection{Final Selection of the Data via the Scattered  Fraction
}\label{subsec:selection}

\begin{figure}
\plotone{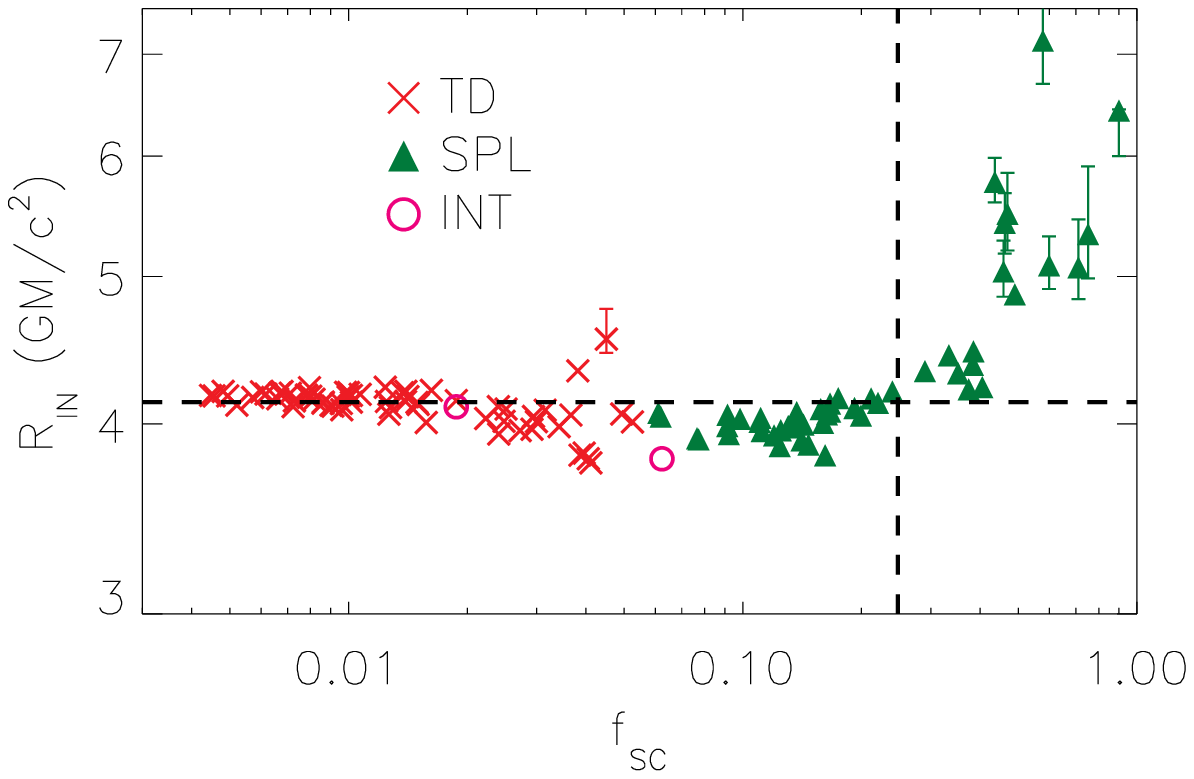} 
\caption{The inner disk radius $\rin$ versus the scattered fraction
  $\fsc$ for H1743.  As indicated in the legend, the symbol types
  denote X-ray state (see \citealt{RM06}).  For $\fsc<0.25$, which is
  our adopted selection criterion, $\rin$ is generally constant; the
  median value for the TD-state data alone is indicated by the dashed
  line.  However, for larger values of $\fsc$, to the right of the
  vertical dashed line, the values of $\rin$ diverge.  Results are
  shown for all 117 spectra with $\rchinu < 2$ and $\rin/\Delta \rin >
  5 $ over the range $L_{D}= 5\%-50\% \;\ledd$ (see
  \S\ref{section:Obs}).  Error bars ($1\sigma$) on $\rin$ that are
  smaller than the plotting symbols have been omitted for clarity.
  Error bars on $\fsc$ are not shown; they are smaller than the
  plotting symbols except for extreme values of $\fsc$ ($<0.02$ and
  $>0.6$).  }
\label{fig:fsc}
\end{figure}

The scattered fraction $\fsc$ sets the strength of the Compton
power-law component relative to the disk.  Figure~\ref{fig:fsc} shows
for H1743 the inner disk radius $\rin$ versus $\fsc$.  For $\fsc <
0.25$ the radius is quite stable and its value for the SPL data is
very nearly the same as for the TD data.  However, at large values of
$\fsc$ the inner disk radius $\rin$ apparently recedes, indicating
that either the model breaks down or a real change takes place in the
disk.  One possible physical explanation was proposed by
\citet{Done_Kubota_2006}, who argue that in regimes of extremely-high
Comptonization an inner disk corona can truncate the disk and increase
$\rin$ by tens of percent, consistent with the high values shown in
Figure~\ref{fig:fsc}.

We have computed and compared plots of $\rin$ versus $\fsc$ for four
BH binaries (H1743, J1550, XTE J1655--40, and LMC X--3) and find that
divergent behavior in their values of $\rin$ sets in for $\fsc \gtrsim
0.2-0.3$ (or $\fsc \gtrsim 0.25-0.4$ for \simpltwob).  Based on a
consideration of these results, we adopt $\fsc < 0.25$ as a
data-selection criterion in this work.  The application of this
criterion leaves a final data sample of 100 spectra for H1743 and 136
for J1550.

\subsection{Comparison with Other Comptonization
  Models}\label{subsec:compare}

Having applied our Comptonized accretion disk model {\sc
  phabs}(\simplb$\otimes$\kerrbbtwob) and obtained final data samples
for H1743 and J1550 (\S\ref{subsec:selection}), we now analyze these
selected data using alternative models for the Compton component.  We
employ (1) \compttb, a widely-used model of Comptonization that
describes the up-scattering of blackbody-like radiation by coronal
electrons \citep{COMPTT}, and (2) the empirical model \powerlawb.  The
full model formulations are respectively {\sc
  phabs}(\kerrbbtwob+\compttb) and {\sc
  phabs}(\kerrbbtwob+\powerlawb).  We now use these models to derive
values of $\rin$ for both sources and compare these results to those
obtained using our model.

\begin{figure*}
\plotone{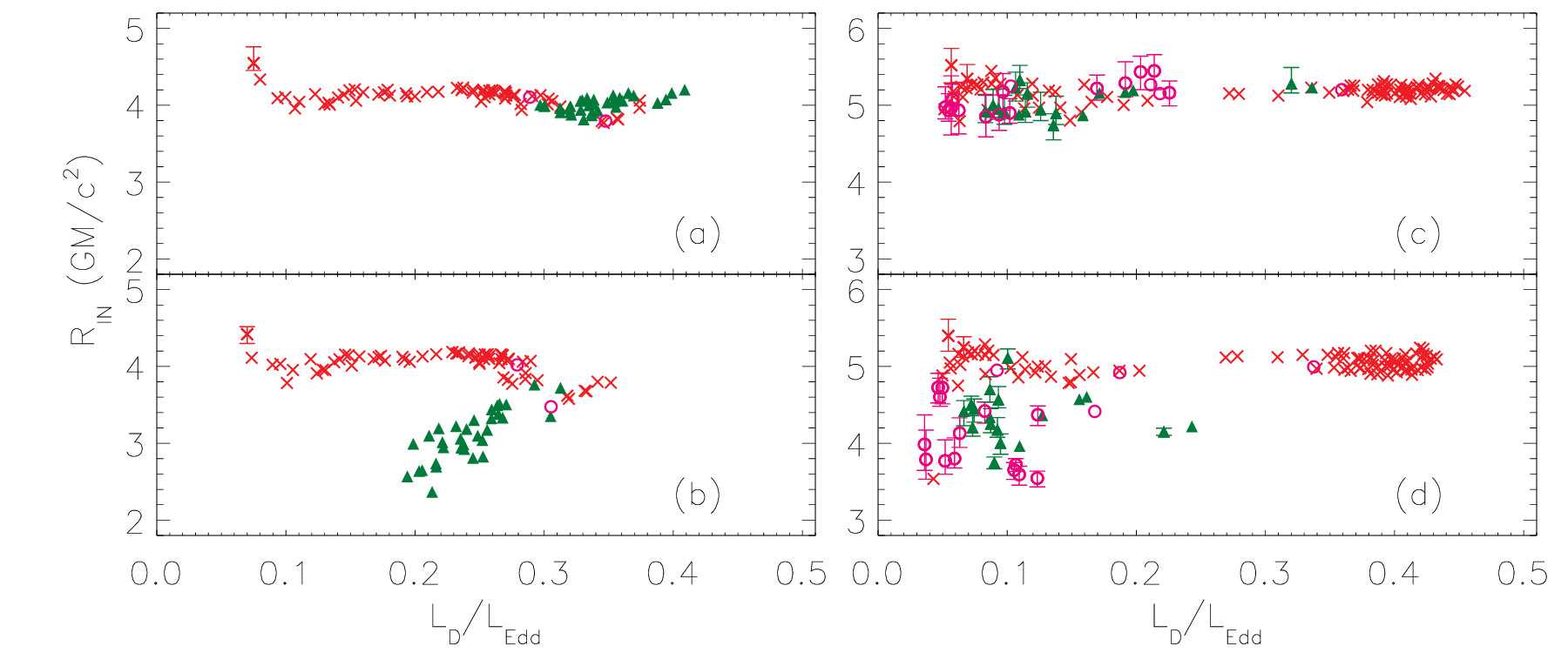}
\caption{The inner disk radius $\rin$ versus the Eddington-scaled disk
  luminosity $L_{D}/\ledd$ for H1743 (left) and J1550 (right).  Symbol
  types are defined in Fig.~\ref{fig:fsc}.  For the upper pair of
  panels the Comptonization model employed is {\sc simpl} and for the
  lower panels it is {\sc powerlaw}.  The data sample considered here
  is that described in \S\ref{subsec:selection}.  For J1550 note in
  panel $c$ the many INT-state data that are brought into agreement
  with the SPL- and TD-state data when applying {\sc simpl}.  Error
  bars are omitted when smaller than the symbols.  }\label{fig:pokerr}
\end{figure*}

Figure \ref{fig:pokerr} shows a side-by-side comparison of H1743 (left
panels) and J1550 (right panels), where $\rin$ is now plotted versus
$L_{D}/\ledd$ (\S\ref{section:Obs}).  The results in the upper pair of
panels were obtained using our self-consistent Comptonization model
{\sc simpl}, while those in the lower panels were obtained using
\powerlawb.  Plainly, for both sources \simpl harmonizes the extreme
discord between the SPL/intermediate (INT) data and the TD data that
results from analyzing these data using \powerlaw
(Fig.\ \ref{fig:pokerr}$b~\&~d$).  The reconciliation achieved using
\simpl (Fig.\ \ref{fig:pokerr}$a~\&~c$) indicates that the inner disk
radii determined in the weakly-Comptonized TD state are very nearly
the same as in the moderately-Comptonized INT and SPL states.  Only
data matching the selection criteria in
\S\S\ref{section:Obs},\ref{subsec:selection} are considered.

Table \ref{tab:rin} provides a summary of the results shown in Figure
\ref{fig:pokerr} and extends the comparison by including results for
    {\sc compTT}.  Qualitatively, the results for both sources are
    very similar; here we comment only on the results for H1743.
    Comparing \simpl with {\sc powerlaw}, we see that for the former
    model $\rin$ is consistent between the TD and SPL states,
    $4.13\pm0.05$ and $4.01\pm0.06$, respectively (values and errors
    here are the median and median absolute deviation).  On the other
    hand, \powerlaw delivers a radius for the SPL state that is
    $\approx 24$\% smaller than for the TD state: $3.10\pm0.24$ versus
    $4.10\pm 0.06$.  While \powerlaw fails dramatically to reconcile
    the TD- and SPL-state data, \comptt provides only a modest
    improvement, giving an $\approx 16$\% smaller value of $\rin$ for
    the SPL state: $3.46\pm0.27$ versus $4.10\pm0.07$.  The failure of
    \comptt and \powerlaw to deliver a constant radius occurs because
    these additive models compete with the disk component for thermal
    flux and because they make no allowance for the flux which the
    disk contributes to the power law.

  \begin{deluxetable}{llcccccccccc} 
  \tabletypesize{\scriptsize} \tablecolumns{12} \tablewidth{0pc}
  \tablecaption{Comparison of $\rin$ Across Spectral States}
  \tablehead{& & & & \multicolumn{3}{c}{$\rin$ (in $\frac{GM}{c^2}$):
      {\sc kerrbb2} used with } \\ \cline{5-7} \\ \colhead{BH} &
    \colhead{State} & \colhead{N} &
    \colhead{${\overline{\fsc}}$\tablenotemark{a}} &
    \colhead{\simplone} & \colhead{\powerlaw} & \colhead{\comptt
      \tablenotemark{b}}} \startdata H1743 & TD & 65 & 0.012 & 4.13
  $\pm$ 0.05 & 4.10 $\pm$ 0.06 & 4.10 $\pm$ 0.07 \\ & INT & 2 & 0.062
  & 3.79 $-$ 4.10 & 3.48 $-$ 4.02 & 3.73 $-$ 4.08 \\ & SPL & 33 &
  0.135 & 4.01 $\pm$ 0.06 & 3.10 $\pm$ 0.24 & 3.46 $\pm$ 0.27
  \\ \tableline J1550 & TD & 100 & 0.016 & 5.20 $\pm$ 0.06 & 5.05
  $\pm$ 0.09 & 5.14 $\pm$ 0.10 \\ & INT & 18 & 0.183 & 5.16 $\pm$ 0.19
  & 4.37 $\pm$ 0.57 & 4.93 $\pm$ 0.20 \\ & SPL & 18 & 0.123 & 5.00
  $\pm$ 0.15 & 4.36 $\pm$ 0.21 & 4.91 $\pm$ 0.26 \\
\enddata

\tablenotetext{a}{Calculated for \simploneb.
For fits using \simpltwob, ${\overline{\fsc}}$ is $\approx$30\%
larger.}  \tablenotetext{b}{Geometry switch set to 1 (slab geometry)
  and redshift to 0.  All other parameters are left free.}

\tablecomments{ The values and errors quoted for $\rin$ are medians
  and median absolute deviations (MADs); we have chosen these
  quantities for their robustness.  For Gaussian-distributed data,
  1$\sigma \approx 1.5$ MAD.  $\rin$ is calculated using the fiducial
  $M$, $i$, and $D$ specified in \S\ref{section:Obs}.}

\label{tab:rin}
\end{deluxetable}


\subsection{Dependence on the Dynamical Model}\label{subsec:dynmod} 

So far, our results are based on the specific and rather arbitrary
dynamical model defined for each source in \S\ref{section:Obs}.  We
now demonstrate that the quality of our results does not depend on the
choice of a particular triplet of $M$, $i$, and $D$.  For H1743 we
analyze the data for six disparate dynamical models chosen as follows:
We fix the mass at $M=10~\msun$ and vary the inclination from
$i=30\degr$ to $i=80\degr$ in $10\degr$ increments, adjusting the
distance in order to maintain the peak disk luminosity at $\lledd
\approx 0.7$; this prescription leaves our selection criteria
(\S\S2,3.1) largely unaffected.
For this demonstration we restrict ourselves to a contiguous set of
pristine data that are free of both edge and line features (see
\citealt{JEM_H1743_2007}).

\begin{figure}
\plotone{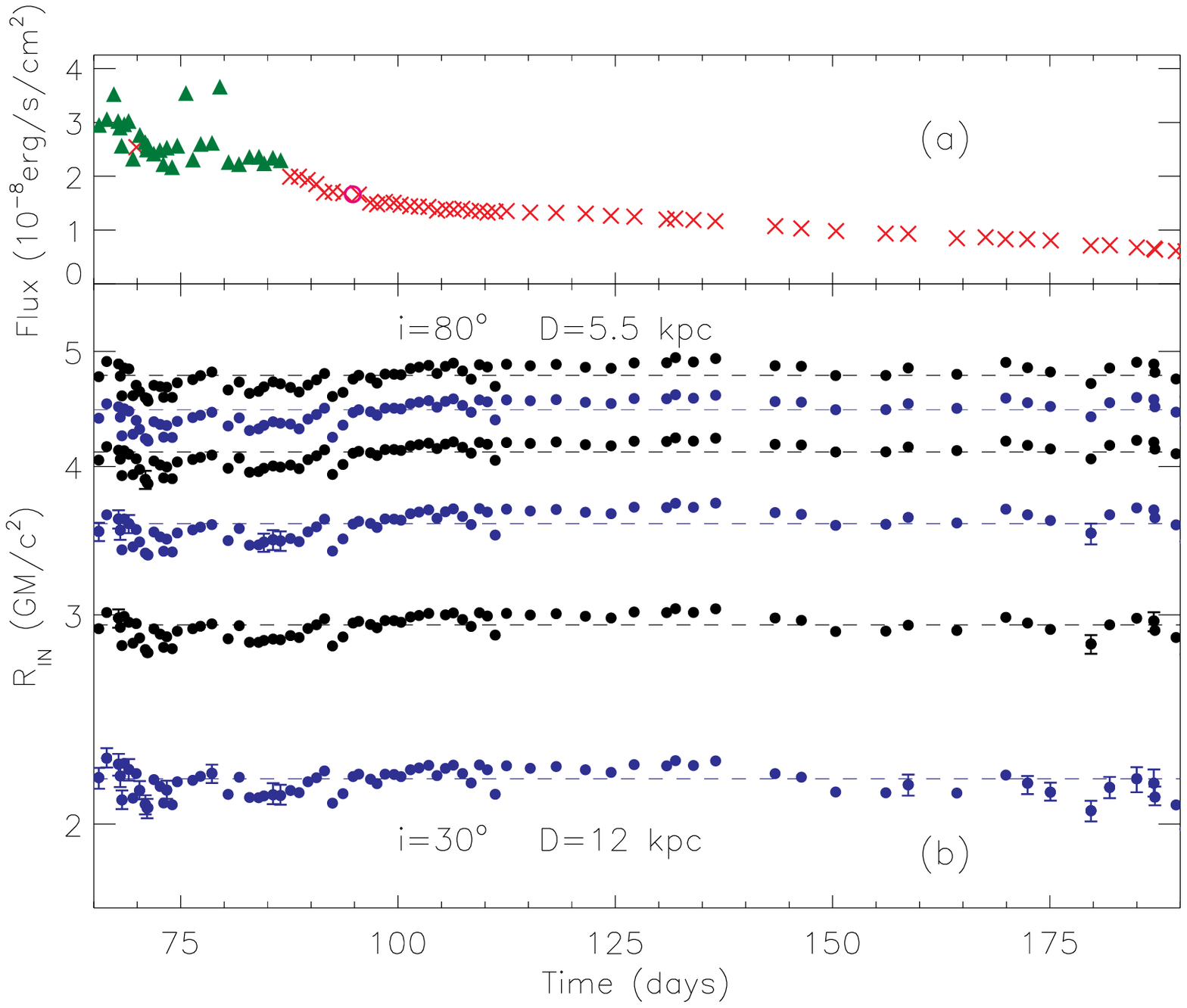}
\caption{($a$) A contiguous 126-day portion of the 225-day {\it RXTE}
  PCA light curve of H1743, which is shown in full in Figure 3$a$ of
  \citet{JEM_H1743_2007}.  The 2--20 keV unabsorbed fluxes were
  obtained by modeling the PCA spectral data.  Time zero is the date
  of discovery of H1743 during its 2003 outburst, which occurred on
  2003 March 21 (MJD 52719).  ($b$) $\rin$ versus time for the six
  models described in the text, shown as alternating black/blue tracks
  for clarity.  The median absolute deviations for the extreme models
  with $i=30^{\circ}$ and $i=80^{\circ}$ are 2.2\% and 1.8\%,
  respectively.  Fluxes for 83 spectra are plotted in panel $a$ and 79
  values of $\rin$ are plotted in panel $b$ (except for $i=40^{\circ}$
  with 78); i.e., four (five for $i=40^{\circ}$) spectra failed to
  meet our selection criteria.
    Error bars are omitted where they are smaller than the
    symbols.}\label{fig:dynmod}
\end{figure}

Figure \ref{fig:dynmod}$a$ shows a portion of the 2003 outburst light
curve of H1743.
Figure \ref{fig:dynmod}$b$ shows corresponding values of $\rin$ versus
time for the six models described above.  We draw the following key
conclusions from Figure \ref{fig:dynmod}: (1) $\rin$ is constant for
each model to within $\approx 2$\% as the source passes from the SPL
state to the TD state, and as the source flux decays by a factor of
$\approx 6$.  We furthermore note that $\rin$ is stable during the two
strong SPL-state flares that occur on days 75.6 and 79.5.  (2) The
character of the small systematic variations that occur in $\rin$
during this entire 4-month period are essentially the same for all six
models.  For completeness, we recomputed all the results shown in
Figure \ref{fig:dynmod}$b$ using first $M=5~\msun$ and then
$M=15~\msun$.  Apart from offsetting the value of $\rin$, the
character of these results is the same, including the level of
scatter, as for the case of $M=10~\msun$.  We conclude that, apart
from setting the median value of $\rin$, the choice of model has no
significant effect on the results presented in
Figure~\ref{fig:dynmod}.

Likewise, for J1550 we analyzed a $\sim 130$-day stretch of data
obtained during the 1998 outburst cycle (MJD 51110 -- 51242;
\citealt{Sobczak_2000}).  We assumed fiducial values of $M$ and $i$
and explored a wide range of distances from $D=3-8~\kpc$.  We obtained
results very similar to those presented for H1743 (Figure
\ref{fig:dynmod}$b$), consistent with an internal scatter of $\approx
2\%$.

\section{Discussion}\label{section:Discussion}

\citet{Kubota_2001} and \citet{Kubota_2004} present the first
self-consistent treatment of disk-dominated accretion at high
luminosity in black hole binaries.  They showed for GRO J1655--40 and
J1550 that what previously had appeared to be anomalous behavior was a
natural result of strong inverse-Compton scattering.  In particular,
they demonstrated that the inner disk radius was stable
when the flux attributed to the power law was properly associated with
the disk.  Their results have been confirmed recently by
\citet{Steiner_simpl} using {\sc simpl} (\S\S1,2).  In this paper, we
provide additional support for the work of Kubota et al., while
supplying in this context the first relativistic analysis of the
accretion disk component.  Both the earlier work by Kubota et al.\ and
this Letter demonstrate that, when modeling Comptonization, a
self-consistent treatment is necessary in order to explain BH behavior
across spectral states.

In all of our earlier work measuring the spins of BHs using {\sc
  kerrbb2}, we have selected data with $\lledd<0.3$, which corresponds
to the thin-disk limit ($H/R \lesssim 0.1$; \citealt{spin_1915}).  In
the present work, the luminosity of J1550 is very uncertain and that
of H1743 is unconstrained.  For this reason, we present a broad range
of luminosities, which likely exceeds the thin-disk limit.  In work
aimed at determining BH spin, when reliable distance estimates and
dynamical data are available, one should apply the aforementioned
luminosity restriction.


In conclusion, we have analyzed a selected sample of $\sim 100$
spectra for each of two bright transient sources using the
self-consistent Comptonization model {\sc simpl} convolved with a
relativistic accretion disk model.  We have thereby shown that the
derived inner disk radii -- or, equivalently, the derived spins of
these BHs -- remain stable to a few percent whether the source is in
the TD state or the more strongly-Comptonized SPL state.  We have
further shown that this stability holds for $\fsc \lesssim 0.25$ and
for a wide range of input model parameters.  We conclude that the
continuum-fitting method of estimating BH spin can be applied to far
more X-ray spectral data and more sources than previously thought
possible.

\acknowledgements

The authors thank Jifeng Liu for valuable discussions.  JFS was
supported by the Smithsonian Institution Endowment Funds and JEM
acknowledges support from NASA grant NNX08AJ55G.  RN acknowledges
support from NASA grant NNX08AH32G and NSF grant AST-0805832. RR
acknowledges partial support from the NASA contract to MIT for support
of \rxte instruments.


\newcounter{BIBcounter}        
\refstepcounter{BIBcounter}    


\begin{thebibliography}{24}
\expandafter\ifx\csname natexlab\endcsname\relax\def\natexlab#1{#1}\fi

\bibitem[{{Arnaud}(1996)}]{XSPEC}
{Arnaud}, K.~A. 1996, in ASP Conf. Ser., Vol. 101, Astronomical Data Analysis
  Software and Systems V, ed. G.~H. {Jacoby} \& J.~{Barnes}, 17

\bibitem[{{Brenneman} \& {Reynolds}(2006)}]{Brenneman_Reynolds}
{Brenneman}, L.~W., \& {Reynolds}, C.~S. 2006, \apj, 652, 1028

\bibitem[{{Davis} \& {Hubeny}(2006)}]{BHSPEC}
{Davis}, S.~W., \& {Hubeny}, I. 2006, \apjs, 164, 530

\bibitem[{{Done} \& {Kubota}(2006)}]{Done_Kubota_2006}
{Done}, C., \& {Kubota}, A. 2006, \mnras, 371, 1216

\bibitem[{{Gou} {et~al.}(2009){Gou}, {McClintock}, {Liu}, {Narayan}, {Steiner},
  {Remillard}, {Orosz}, \& {Davis}}]{Gou_2009}
{Gou}, L., {McClintock}, J.~E., {Liu}, J., {Narayan}, R., {Steiner}, J.~F.,
  {Remillard}, R.~A., {Orosz}, J.~A., \& {Davis}, S.~W. 2009, ApJ, submitted
  (arXiv:0901.0920v1 [astro-ph.HE])



\bibitem[{Jahoda} {et~al.}(2006)]{RXTE_cal} {Jahoda}, K., {Markwardt}, 
C.~B., {Radeva}, Y., {Rots}, A.~H., {Stark}, M.~J., {Swank}, J.~H., {Strohmayer}, 
T.~E., \& {Zhang}, W.  2006, \apjs, 163, 401 



\bibitem[{{Kubota} \& {Makishima}(2004)}]{Kubota_2004}
{Kubota}, A., \& {Makishima}, K. 2004, \apj, 601, 428

\bibitem[{{Kubota} {et~al.}(2001){Kubota}, {Makishima}, \&
  {Ebisawa}}]{Kubota_2001}
{Kubota}, A., {Makishima}, K., \& {Ebisawa}, K. 2001, \apjl, 560, L147

\bibitem[{{Li} {et~al.}(2005){Li}, {Zimmerman}, {Narayan}, \&
  {McClintock}}]{KERRBB}
{Li}, L.-X., {Zimmerman}, E.~R., {Narayan}, R., \& {McClintock}, J.~E. 2005,
  \apjs, 157, 335

\bibitem[{{Liu} {et~al.}(2008){Liu}, {McClintock}, {Narayan}, {Davis}, \&
  {Orosz}}]{spin_m33}
{Liu}, J., {McClintock}, J.~E., {Narayan}, R., {Davis}, S.~W., \& {Orosz},
  J.~A. 2008, \apjl, 679, L37

\bibitem[{{McClintock} {et~al.}(2009){McClintock}, {Remillard}, {Rupen},
  {Torres}, {Steeghs}, {Levine}, \& {Orosz}}]{JEM_H1743_2007}
{McClintock}, J.~E., {Remillard}, R.~A., {Rupen}, M.~P., {Torres}, M.~A.~P.,
  {Steeghs}, D., {Levine}, A.~M., \& {Orosz}, J.~A. 2009, \apj, 698, 1398 


\bibitem[{{McClintock} {et~al.}(2006){McClintock}, {Shafee}, {Narayan},
  {Remillard}, {Davis}, \& {Li}}]{spin_1915}
{McClintock}, J.~E., {Shafee}, R., {Narayan}, R., {Remillard}, R.~A., {Davis},
  S.~W., \& {Li}, L.-X. 2006, \apj, 652, 518

\bibitem[{{Miller} {et~al.}(2003){Miller}, {Marshall}, {Wijnands}, {Di Matteo},
  {Fox}, {Kommers}, {Pooley}, {Belloni}, {Casares}, {Charles}, {Fabian}, {van
  der Klis}, \& {Lewin}}]{Miller_2003_j1550}
{Miller}, J.~M., {et~al.} 2003, \mnras, 338, 7

\bibitem[{{Miller} {et~al.}(2009){Miller}, {Reynolds}, {Fabian}, {Miniutti}, \&
  {Gallo}}]{Miller_2009}
{Miller}, J.~M., {Reynolds}, C.~S., {Fabian}, A.~C., {Miniutti}, G., \&
  {Gallo}, L.~C. 2009, ApJ, in press (arXiv:0902.2840v1 [astro-ph.HE])

\bibitem[{{Miniutti} {et~al.}(2007){Miniutti}, {Fabian}, {Anabuki}, {Crummy},
  {Fukazawa}, {Gallo}, {Haba}, {Hayashida}, {Holt}, {Kunieda}, {Larsson},
  {Markowitz}, {Matsumoto}, {Ohno}, {Reeves}, {Takahashi}, {Tanaka},
  {Terashima}, {Torii}, {Ueda}, {Ushio}, {Watanabe}, {Yamauchi}, \&
  {Yaqoob}}]{Miniutti_2007}
{Miniutti}, G., {et~al.} 2007, \pasj, 59, 315

\bibitem[{{Orosz} {et~al.}(2002){Orosz}, {Groot}, {van der Klis}, {McClintock},
  {Garcia}, {Zhao}, {Jain}, {Bailyn}, \& {Remillard}}]{Orosz_2002}
{Orosz}, J.~A., {et~al.} 2002, \apj, 568, 845

\bibitem[{{Remillard} \& {McClintock}(2006)}]{RM06}
{Remillard}, R.~A., \& {McClintock}, J.~E. 2006, \araa, 44, 49

\bibitem[{{Shafee} {et~al.}(2006){Shafee}, {McClintock}, {Narayan}, {Davis},
  {Li}, \& {Remillard}}]{Shafee_spin}
{Shafee}, R., {McClintock}, J.~E., {Narayan}, R., {Davis}, S.~W., {Li}, L.-X.,
  \& {Remillard}, R.~A. 2006, \apjl, 636, L113

\bibitem[{{Shapiro} \& {Teukolsky}(1983)}]{Shapiro_Teukolsky}
{Shapiro}, S.~L., \& {Teukolsky}, S.~A. 1983, {Black Holes, White Dwarfs, and
  Neutron Stars} (New York: Wiley-Interscience)


\bibitem[Sobczak et al.(2000)]{Sobczak_2000} Sobczak, G.~J., 
McClintock, J.~E., Remillard, R.~A., Cui, W., Levine, A.~M., Morgan, E.~H., 
Orosz, J.~A., \& Bailyn, C.~D.\ 2000, \apj, 544, 993 



\bibitem[{{Steiner} {et~al.}(2009){Steiner}, {Narayan}, {McClintock}, \&
  {Ebisawa}}]{Steiner_simpl}
{Steiner}, J.~F., {Narayan}, R., {McClintock}, J.~E., \& {Ebisawa}, K. 2009,
  PASP, submitted (arXiv:0810.1758v2 [astro-ph])

\bibitem[{{Swank}(1999)}]{RXTE}
{Swank}, J.~H. 1999, Nuclear Physics B Proc. Suppl., 69, 12

\bibitem[Tanaka et al.(1995)]{Tanaka_1995} Tanaka, Y., et al.\ 
1995, \nat, 375, 659 


\bibitem[{{Titarchuk}(1994)}]{COMPTT}
{Titarchuk}, L. 1994, \apj, 434, 570

\bibitem[{{Toor} \& {Seward}(1974)}]{Toor_seward}
{Toor}, A., \& {Seward}, F.~D. 1974, \aj, 79, 995

\bibitem[{{Zhang} {et~al.}(1997){Zhang}, {Cui}, \& {Chen}}]{Zhang}
{Zhang}, S.~N., {Cui}, W., \& {Chen}, W. 1997, \apjl, 482, L155


\end{thebibliography}

\mbox{~}

\end{document}